# Broad band invisibility cloak made of normal dielectric multilayer

Xiaofei Xu, Yijun Feng[*], Shuai Xiong, Jinlong Fan, Jun-Ming Zhao, and Tian Jiang

Department of Electronic Engineering, School of Electronic Science and Engineering, Nanjing

Univerisity, Nanjing, 210093, China

(Revised: Aug. 30, 2011)

**Abstract**

We present the design, fabrication and performance test of a quasi three-dimensional carpet cloak made of normal dielectric in the microwave regime. Taking advantage of a simple linear coordinate transformation we design a carpet cloak with homogeneous anisotropic medium and then practically realize the device with multilayer of alternating normal dielectric slabs based on the effective medium theory. As a proof-of-concept example, we fabricate the carpet cloak with multilayer of FR4 dielectric slabs with air spacing. The performance of the fabricated design is verified through full-wave numerical simulation and measurement of the far-field scattering electromagnetic waves in a microwave anechoic chamber. Experimental results have demonstrated pronounced cloaking effect in a very broad band from 8 GHz to 18 GHz (whole X and Ku band) due to the low loss, non-dispersive feature of the multilayer dielectric structure.

---

[*] Author to whom correspondence should be addressed. Electronic mail: yjfeng@nju.edu.cn

Recent progress in transformation optics (TO) [1-2] provides us a precise design tool of engineered metamaterials that could be applied to the manipulation of electromagnetic (EM) waves, such as the invisibility cloaking and other interesting applications [3-6]. Perfect invisibility cloak could deflect incident EM wave/light ray around an inner core region, and thus the objects inside the core become transparent to the outside EM detection [1]. However, the perfect invisibility cloak suffers from singular material properties and dispersions, which limits its realization and application. To avoid these limitations, several reduced invisibility cloaks with simplified constitutive parameters have been proposed and realized with the resonant metamaterial structures [7-9]. Recently, to overcome the narrow-band operation of the resonant metamaterials, a special cloaking scheme, called the 'carpet cloak' which is based on quasi-conformal coordinate transformation, has been proposed [10] and studied [11-13]. These carpet cloaks work at a rather broad band and could conceal objects underneath a reflecting bulging surface, imitating the reflection of a flat surface.

Most carpet cloaks are generated from quasi-conformal transformation, which leads to minimal anisotropy and facilitates the fabrications with metamaterials [14-20]. However, the material properties retrieved from this procedure are highly inhomogeneous, which in the microwave regime usually rely on large number of precisely defined sub-wavelength artificial structures [14, 15], or require accurately defined nano-scale patterns made by sophisticated time-consuming nanofabrication processes in the optical regime [16-20]. Another drawback, as recently suggested in a theoretical study [21], is that the isotropy cloak structure derived from the quasi-conformal transformation will lead to lateral shift of the scattering fields, making the carpet cloak still visible to near field detection. Another attempt to achieve carpet cloak is to use simple linear coordinate transformation which results in cloak with anisotropic but homogeneous material parameters [22, 23]. This proposal has been applied in the infrared frequency to realize two-dimensional cloak with silicon grating structures embedded in a silicon on insulator wafer [24, 25]. Recently, this proposal has also been used to achieve macroscopic invisibility cloak based on the natural birefringent crystal calcite and the cloak performance has been experimentally demonstrated for a specific polarization at the visible frequencies [26, 27]. However, since the natural transparent birefringent crystals are quite limited and their anisotropy can not be engineered as desired, the design flexibility is restricted for this type of cloak.

In this letter, we report an alternative way to realize the required anisotropic homogeneous carpet cloak through normal dielectric multilayer. Based on the effective medium theory (EMT), an anisotropic homogeneous medium can be realized by alternating normal dielectric layers, and its anisotropy can be engineered flexibly through the alignment angle and thickness ratio of the multilayer [28, 29]. We design, fabricate and test a quasi three-dimensional (3D) broad band carpet cloak made of multilayer of normal dielectric slabs with air spacing in the microwave regime. The performance of the carpet cloak through out the whole X, and Ku band (from 8 GHz to 18 GHz) is validated through full wave EM simulation based on finite element method and measurement of the EM scattering fields in a microwave anechoic chamber.

We begin with a straightforward linear coordinate transformation discussed in [22-24]. Fig. 1(a) indicates schematically the coordinate transformation under which a virtual space ($x$, $y$, $z$) with a triangular cross-section in the $x$-$y$ plane is squeezed into a region with a quadrilateral cross-section (light blue region) in the physical space ($x'$, $y'$, $z'$), leaving the lower triangle part as the cloaked region. The whole carpet cloak is a 3D structure stretched infinitely and uniformly along the $z$ axis. Supposing the outer surface of the carpet cloak or the cloaked region has a the slope of $k_1$, or $k_2$, respectively, the transformation can be defined as $x' = x$, $y' = (k_1 - k_2) y / k_1 + k_2 (a + x)$, $z' = z$. We could then obtain the material parameters for the left block of the carpet cloak with the standard TO procedures, similar to that described in Eq. (1) in [24]. We further consider a transverse-magnetic (TM) field polarization for the incident EM wave (the magnetic field along the $z$ axis), and as far as the trajectory of EM wave is concerned, the material parameters of the cloak can be simplified to a nonmagnetic form with $\mu = 1$, and

$$\bar{\bar{\varepsilon}}' = \varepsilon_r \begin{pmatrix} [k_1/(k_1 - k_2)]^2 & k_2[k_1/(k_1 - k_2)]^2 & 0 \\ k_2[k_1/(k_1 - k_2)]^2 & k_2^2[k_1/(k_1 - k_2)]^2 + 1 & 0 \\ 0 & 0 & \varepsilon_z \end{pmatrix}, \quad (1)$$

where $\bar{\bar{\varepsilon}}'$ represents the permittivity tensor for the cloak in the physical space, while $\varepsilon_r$ represents the relative permittivity of the background medium in the virtual space, respectively. Here $\varepsilon_z$ can be an arbitrary value for the TM polarization.

As a direct consequence of the simple linear coordinate transformation, we find that the material parameters described in Eq. (1) only require a spatially invariant permittivity tensor, and can be easily realized through a certain birefringent dielectric with its optical axis rotated by a certain angle $\theta$ with the $z$-axis. The principal values of the permittivity tensor $\varepsilon_x^c$, $\varepsilon_y^c$ and $\varepsilon_z^c$ can be determined through diagonalization of the parameters in Eq. (1). For the TM polarization $\varepsilon_z^c$ is arbitrary and can be assumed as $\varepsilon_z^c = \varepsilon_x^c$. Therefore the resulted cloak can be realized with a birefringent dielectric crystal described by $\varepsilon_x^c$ and $\varepsilon_y^c$. Unlike the previous works that use natural birefringent crystal calcite to design the cloak [26, 27], here we employ a more flexible way to realize the birefringent dielectric through multilayer of alternating dielectric 1 and dielectric 2 (indicated schematically in Fig. 1(b)) based on the EMT [28. 29]. The required dielectric materials in the multilayer are determined through

$$\varepsilon_x^c = \varepsilon_z^c = \frac{\varepsilon_1 + \eta \varepsilon_2}{1+\eta}, \quad \varepsilon_y^c = \frac{(1+\eta)\varepsilon_1 \varepsilon_2}{\eta \varepsilon_1 + \varepsilon_2} \quad , \qquad (2)$$

where $\varepsilon_1$ and $\varepsilon_2$ are the permittivity and $\eta$ is the thickness ratio of the two isotropic dielectrics, respectively. We simply let $\eta = 1$ for the consideration of achieving maximum anisotropy. This multilayer approach allows us more freedom to design the carpet cloak with desired shape (different slope $k_1$, or $k_2$) and material parameters.

As a proof-of-concept example of our proposal, we have designed a carpet cloak at microwave regime. Assuming $k_1 = 1$, $k_2 = 0.17$, $a = 100$ mm, and the cloak is immersed in a background medium of $\varepsilon_r = 1.75$. From the above description, the relative permittivity of the dielectric 1 and 2 composing the carpet cloak is determined as 4.4 and 1.0, respectively, and the rotation angle $\theta$ is about 25.8° with respect to $x$-axis. Considering the practical realization, we replace the background medium with free space, which in principle will cause slight reflections at the boundary between the cloak and the background medium due to wave impedance mismatch, but the wave trajectory inside the cloak and thus the cloaking effect will keep unchanged.

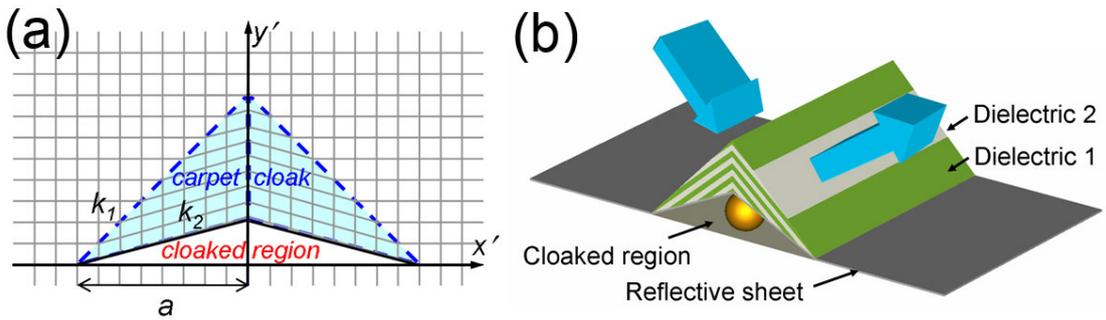

Fig. 1. (Color online) (a) Scheme of the coordinate transformation and the resulted carpet cloak. (b) Realization of the proposed 3D carpet cloak with multilayer of alternating dielectric 1 and 2. The cloak is covered on a bumped perfect conducting surface enabling a cloaked region underneath.

To verify the design, we study how the EM waves interact with the carpet cloak with full-wave numerical calculations based on the finite element method (Comsol Multiphysics). Suppose a Gauss beam with TM polarization incidents along an azimuth angle of $\varphi = 135°$ with respect to $x$-axis, which encounters a diffuse reflection by a reflective bump structure. Fig. 2(a) illustrates the near field distributions of the magnetic field, from which we could easily "detect" the bump structure from the irregular scattering. However, if we consider a background medium with permittivity of 1.75, and cover the bump with an anisotropic carpet cloak determined from Eq. (1), the bump becomes invisible, with a near field distribution of specular reflection mimicking that of a flat reflective sheet (shown in Fig. 2(b)). Consider the practical realization, we remove the background medium with free space, and construct the cloak with multilayer of alternating dielectrics as proposed before. FR4 dielectric (usually used in the print circuit board technology) with permittivity of 4.4 and loss tangent of 0.025 has been used for dielectric 1, while air spacing for dielectric 2. Both have a thickness of 1.5 mm. Fig. 2(c) illustrates the near field distributions which represents a specular reflection similar to that of Fig. 2(b) indicating that the reduced multilayer cloak design still has good performance.

We have also calculated the far-field pattern of the scattering field through the near-to-far-field extrapolation algorithm, and compared in Fig. 2(d). Both the anisotropic and the multilayer cloaks result in confined far-field scattering along the specular direction ($\varphi = 45°$) similar to that from a flat reflective sheet, while the bare bump results in irregular scattering fields with two lobes around $\varphi = 28°$ and $\varphi = 68°$. The slight backward scattering by the anisotropic or the multilayer cloak is due to the nonmagnetic reduction and removal of the background medium resulting in small impedance mismatching.

We fabricate a quasi 3D carpet cloak composed of multilayer of FR4 slab and air spacing as depicted in Fig. 3(a). The cloak is aimed to work in the X and Ku band, covering frequency from 8 GHz to 18 GHz. The center frequency corresponds to a free space wavelength of about 23 mm. The whole structure is about 300 mm long (about 13 times of the wavelength) in the $z$ direction, 200 mm wide in the $x$ direction and 100 mm high, creating a cloak region about 200 mm wide and 17 mm high in the center. To satisfy the effective medium theory described in Eq. (2), either the FR4 slab or air spacing has a thickness of 1.5mm, about 1/15 of the center free space wavelength. The carpet cloak covers a reflective bump (copper board) on an aluminum perfect conducting board with the size of 300 mm × 300 mm.

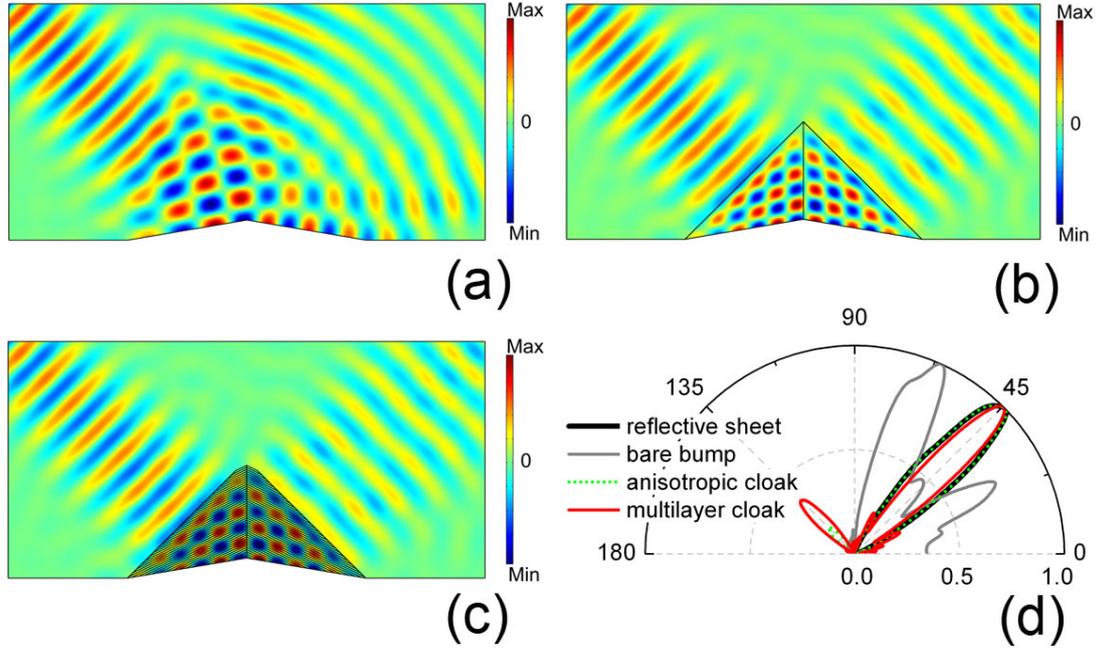

Fig. 2. (Color online) Full-wave simulations of the magnetic field distributions of incident EM beam interacting with (a) bare conducting bump; (b) bump covered by carpet cloak with a background dielectric with permittivity of 1.75; (c) bump covered by carpet cloak composed of multilayer of FR4 slabs with air spacing. (d) The far field distributions of the scattered EM waves for the different cases indicated by the legends.

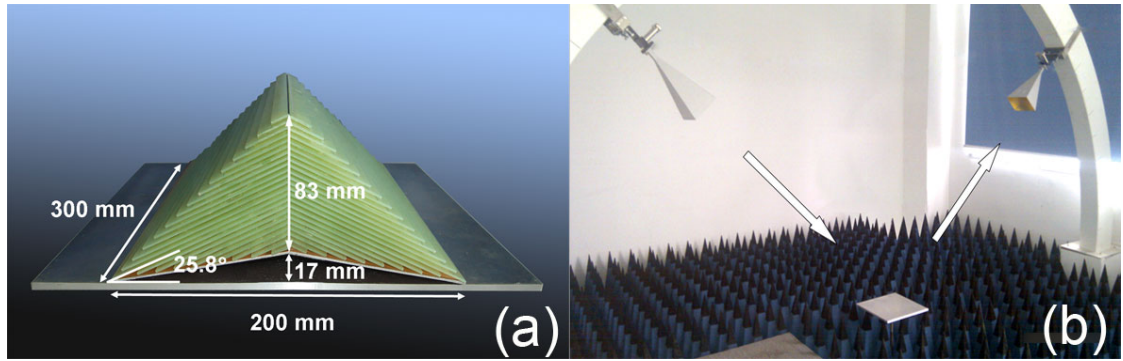

Fig. 3. (Color online) Photographs of (a) the fabricated multilayer cloak, and (b) the measurement system.

The performance of the multilayer cloak is tested in a microwave anechoic chamber. Two standard rectangular horn antennas are used as the transmitter and receiver and linked to a vector network analyzer to measure the scattering field. Both horns can be moved along an arc rail to change their azimuth angles as shown in Fig. 3(b). To obtain a quasi plane wave of TM polarization, we mainly take the advantage of the far field phenomenon of the horns. We fix the incident wave at an azimuth angle of 135°, and record the scattering field as a function of the azimuth angle $\varphi$ from 10° to 80° through out the whole X and Ku band. The scattering field amplitude has been normalized to that of the specular reflection by a flat conducting plate of 300 mm × 300 mm, and the results are compared in Fig. 4 for three different cases: flat conducting plate, conducting bump, and the bump covered by the multilayer cloak. As shown in Fig. 4(a) and (b), the flat conducting sheet produce an idea mirror-reflection with single peak scattering at the specular direction ($\varphi$ =45°). However, the bumped conducting sheet generates irregular scattering with two main lobes roughly around $\varphi$ = 30° and $\varphi$ = 70°, as shown in Fig. 4(c) and (d), which agrees roughly with the simulation in Fig. 2(d). When the bump is covered with the multilayer cloak, the mirror-reflection is restored with pronounced single peak around $\varphi$ =45° as shown in Fig. 4(e) and (f), imitating that of the flat reflective sheet without any bump. The experiment results have obviously validated the design procedure of realizing the carpet cloak through multilayer of normal dielectrics. The cloaking performance has been verified at different incident angles numerically and experimentally, as the cloak is designed based on transformation optics and should work for all incident angles. The cloak also demonstrates a very broad band performance through out the whole X and Ku band, which is due to the non-dispersive and low loss feature of the multilayer structure.

In conclusion, we have reported the design, fabrication and test of a quasi 3D carpet cloak composed of multilayer of alternating normal dielectric slabs at microwave regime. Through a simple linear coordinate transformation, the carpet cloak can be designed with homogeneous birefringent dielectric medium. Then a practical example of the cloak device has been constructed by realizing the required birefringent dielectric with multilayer of stacked FR4 dielectric slabs and air spacing based on the EMT. The effectiveness of the device has been verified through full-wave simulation and validated by measurement of the far-field scattering pattern on the fabricated sample, which demonstrated the broad band cloaking effect through out the whole X and Ku band. The proposed design procedure through multilayer of normal dielectrics provides us a cost effective and easy-to-fabricate way to obtain EM cloaking device, and can be further scale to terahertz or optical frequency regime through conventional multilayer thin film technology. We also believe the multilayer approach allows us more freedom in realization of dielectric medium with desired anisotropy of material parameters, and thus can be applied to experimental demonstrations of other transformation optics devices.

This work is supported by the National Nature Science Foundation (60990322, 60990320, 60801001 and 60671002), the Fundamental Research Funds for the Central Universities (1111021001,

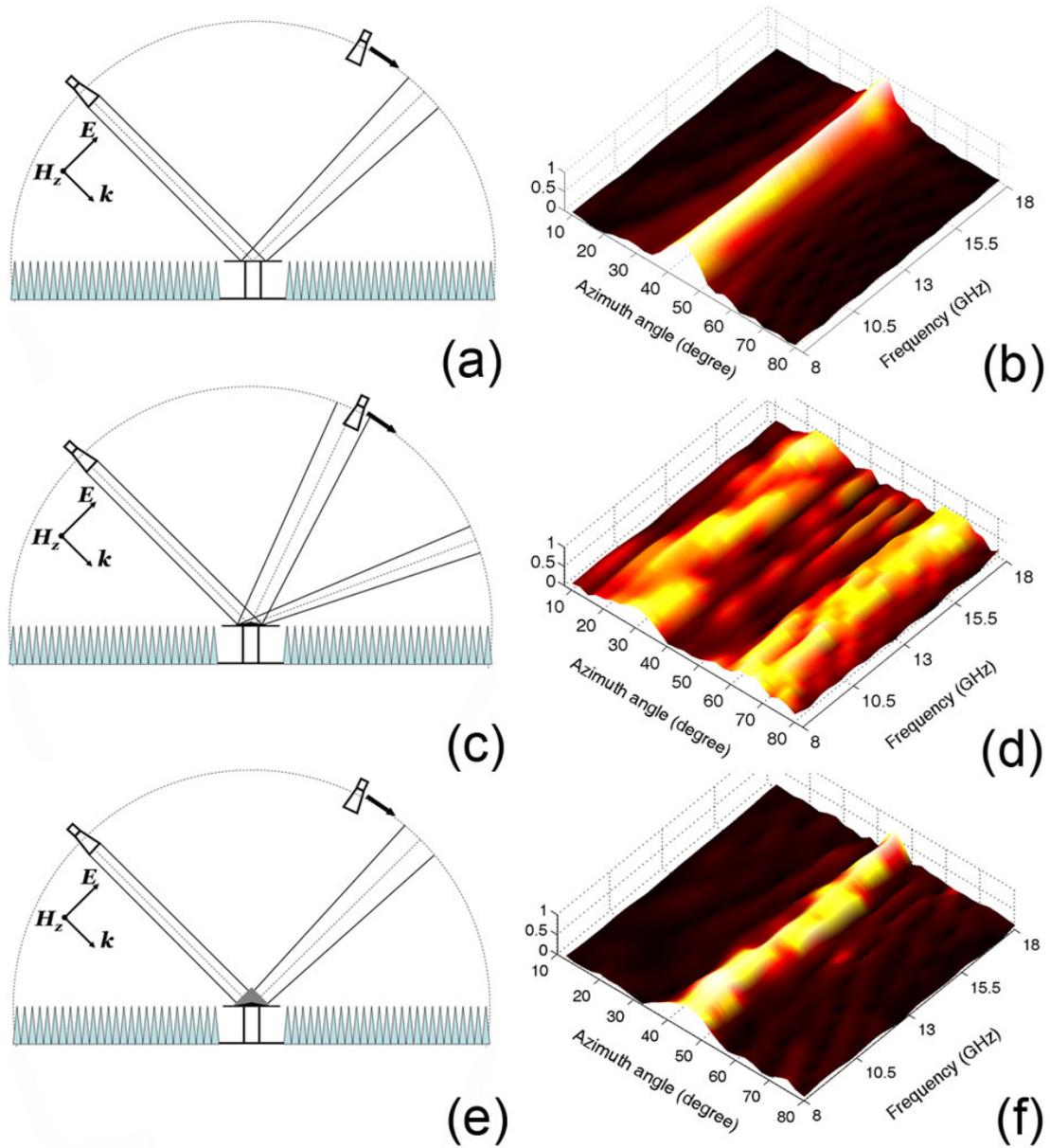

Fig. 4. (Color online) Experimental scheme and the measured results of the far-field scattering EM field magnitude as a function of azimuth angle and frequency. (a)-(b) the flat conducting sheet; (c)-(d) the bare conducing bump; and (e)-(f) bump covered by the multilayer carpet cloak.

**References**

1. J. B. Pendry, D. Schurig, and D. R. Smith, Science **312**, 1780 (2006).

2. U. Leonhardt, Science **312**, 1777 (2006).

3. H.Y. Chen and C. T. Chan, Appl. Phys. Lett. **90**, 241105 (2007).

4. M. Rahm, S. A. Cummer, D. Schurig, J. B. Pendry, and D. R. Smith, Phys. Rev. Lett. **100**, 063903 (2008).

5. X. Xu, Y. Feng, and T. Jiang, New J. Phys. **10**, 115027 (2008).

6. N. Kundtz, and D. R. Smith, Nature Mater. **9**, 129 (2010).

7. D. Schurig, J. J. Mock, B. J. Justice, S. A. Cummer, J. B. Pendry, A. F. Starr, and D. R. Smith, Science **314**, 977 (2006).

8. W. Cai, U. K. Chettiar, A. V. Kildishev, and V. M. Shalaev, Nat. Photonics **1**, 224 (2007).

9. Y. Huang, Y. Feng, and T. Jiang, Opt. Express **15**, 11133 (2007).

10. J. Li and J. B. Pendry, Phys. Rev. Lett. **101**, 203901 (2008).

11. E. Kallos, C. Argyropoulos, and Y. Hao, Phys. Rev. A **79**, 063825 (2009).

12. J.C. Halimeh, T. Ergin, J. Mueller, N. Stenger, and M. Wegener, Opt. Express **17**, 19328 (2009).

13. P. Zhang, M. Lobet, and S. He, Opt. Express **18**, 18158 (2010).

14. R. Liu, C. Ji, J. J. Mock, J. Y. Chin, T. J. Cui, and D. R. Smith, Science **323**, 366 (2009).

15. H. F. Ma, and T. J. Cui, Nat Commun. **1**, 1 (2010).

16. J. Valentine, J. Li, T. Zentgraf, G. Bartal, and X. Zhang, Nat. Mater. **8**, 568 (2009).

17. L. H. Gabrielli, J. Cardenas, C. B. Poitras, and M. Lipson, Nat. Photonics **3**, 461 (2009).

18. T. Ergin, N. Stenger, P. Brenner, J. B. Pendry, and M. Wegener, Science **328**, 337 (2010).

19. M. Gharghi, C. Gladden, T. Zentgraf, Y. Liu, X. Yin, J. Valentine, and X. Zhang, Nano Lett. **11**, 2825 (2011).

20. J. Fischer, T. Ergin, and M. Wegener, Opt. Lett. **36**, 2059 (2011).

21. B. Zhang, T. Chan, and B.-I. Wu, Phys. Rev. Lett. **104**, 233903 (2010).

22. S. Xi, H. Chen, B. Wu, and J. A. Kong, IEEE Microw. Wireless Compon. Lett. **19**, 131 (2009).

23. Y. Luo, J. Zhang, H. Chen, L. Ran, B.-I. Wu, and J. A. Kong, IEEE Trans. Antenn. Propag. **57**, 3926 (2009).

24. X. Xu, Y. Feng, Y. Hao, J. Zhao, and T. Jiang, Appl. Phys. Lett. **95**, 184102 (2009).

25. J. Zhang, L. Liu, Y. Luo, S. Zhang, and N. A. Mortensen, Opt. Express **19**, 8625 (2011).

26. X. Chen, Y. Luo, J. Zhang, K. Jiang, J. B. Pendry, and S. Zhang, Nat. Commun. **2**, 176 (2011).

27. B. Zhang, Y. Luo, X. Liu, and G. Barbastathis, Phys. Rev. Lett. **106**, 033901 (2011).

28. B. Wood, J. B. Pendry, and D. P. Tsai, Phys. Rev. B **74**, 115116 (2006).

29. X. Xu, Y. Feng, Z. Yu, T. Jiang, and J. Zhao, Opt. Express **18**, 24477 (2010).